\documentclass[aps,preprint]{revtex4}
\usepackage{graphicx}
\usepackage[scanall]{psfrag}
\usepackage{amssymb}

\begin{document}

%
%
%
%
\def\oti{{\otimes}}
\def\lb{ \left[ }
\def\rb{ \right]  }
\def\tilde{\widetilde}
\def\bar{\overline}
\def\hat{\widehat}
\def\*{\star}
\def\[{\left[}
\def\]{\right]}
\def\({\left(}		\def\BL{\Bigr(}
\def\){\right)}		\def\BR{\Bigr)}
	\def\BBL{\lb}
	\def\BBR{\rb}
%
%
\def\zb{{\bar{z} }}
\def\zbar{{\bar{z} }}
\def\frac#1#2{{#1 \over #2}}
\def\inv#1{{1 \over #1}}
\def\half{{1 \over 2}}
\def\d{\partial}
\def\der#1{{\partial \over \partial #1}}
\def\dd#1#2{{\partial #1 \over \partial #2}}
\def\vev#1{\langle #1 \rangle}
\def\ket#1{ | #1 \rangle}
\def\rvac{\hbox{$\vert 0\rangle$}}
\def\lvac{\hbox{$\langle 0 \vert $}}
\def\2pi{\hbox{$2\pi i$}}
\def\e#1{{\rm e}^{^{\textstyle #1}}}
\def\grad#1{\,\nabla\!_{{#1}}\,}
\def\dsl{\raise.15ex\hbox{/}\kern-.57em\partial}
\def\Dsl{\,\raise.15ex\hbox{/}\mkern-.13.5mu D}
%
%
\def\ga{\gamma}		\def\Ga{\Gamma}
\def\be{\beta}
\def\al{\alpha}
\def\ep{\epsilon}
\def\vep{\varepsilon}
\def\la{\lambda}	\def\La{\Lambda}
\def\de{\delta}		\def\De{\Delta}
\def\om{\omega}		\def\Om{\Omega}
\def\sig{\sigma}	\def\Sig{\Sigma}
\def\vphi{\varphi}

%
%
\def\CA{{\cal A}}	\def\CB{{\cal B}}	\def\CC{{\cal C}}
\def\CD{{\cal D}}	\def\CE{{\cal E}}	\def\CF{{\cal F}}
\def\CG{{\cal G}}	\def\CH{{\cal H}}	\def\CI{{\cal J}}
\def\CJ{{\cal J}}	\def\CK{{\cal K}}	\def\CL{{\cal L}}
\def\CM{{\cal M}}	\def\CN{{\cal N}}	\def\CO{{\cal O}}
\def\CP{{\cal P}}	\def\CQ{{\cal Q}}	\def\CR{{\cal R}}
\def\CS{{\cal S}}	\def\CT{{\cal T}}	\def\CU{{\cal U}}
\def\CV{{\cal V}}	\def\CW{{\cal W}}	\def\CX{{\cal X}}
\def\CY{{\cal Y}}	\def\CZ{{\cal Z}}

\def\rvac{\hbox{$\vert 0\rangle$}}
\def\lvac{\hbox{$\langle 0 \vert $}}
\def\comm#1#2{ \BBL\ #1\ ,\ #2 \BBR }
\def\2pi{\hbox{$2\pi i$}}
\def\e#1{{\rm e}^{^{\textstyle #1}}}
\def\grad#1{\,\nabla\!_{{#1}}\,}
\def\dsl{\raise.15ex\hbox{/}\kern-.57em\partial}
\def\Dsl{\,\raise.15ex\hbox{/}\mkern-.13.5mu D}
%
%
%
\font\numbers=cmss12
\font\upright=cmu10 scaled\magstep1
\def\stroke{\vrule height8pt width0.4pt depth-0.1pt}
\def\topfleck{\vrule height8pt width0.5pt depth-5.9pt}
\def\botfleck{\vrule height2pt width0.5pt depth0.1pt}
\def\Zmath{\vcenter{\hbox{\numbers\rlap{\rlap{Z}\kern
0.8pt\topfleck}\kern 2.2pt
                   \rlap Z\kern 6pt\botfleck\kern 1pt}}}
\def\Qmath{\vcenter{\hbox{\upright\rlap{\rlap{Q}\kern
                   3.8pt\stroke}\phantom{Q}}}}
\def\Nmath{\vcenter{\hbox{\upright\rlap{I}\kern 1.7pt N}}}
\def\Cmath{\vcenter{\hbox{\upright\rlap{\rlap{C}\kern
                   3.8pt\stroke}\phantom{C}}}}
\def\Rmath{\vcenter{\hbox{\upright\rlap{I}\kern 1.7pt R}}}
\def\Z{\ifmmode\Zmath\else$\Zmath$\fi}
\def\Q{\ifmmode\Qmath\else$\Qmath$\fi}
\def\N{\ifmmode\Nmath\else$\Nmath$\fi}
\def\C{\ifmmode\Cmath\else$\Cmath$\fi}
\def\R{\ifmmode\Rmath\else$\Rmath$\fi}

\def\barray{\begin{eqnarray}}
\def\earray{\end{eqnarray}}
\def\beq{\begin{equation}}
\def\eeq{\end{equation}}

\def\no{\noindent}
\def\chidag{\chi^\dagger}
\def\Det{{\rm Det}}
\def\Tr{{\rm Tr}}
\def\det{{\rm det}}
\def\tr{{\rm tr}}
\def\Upp{ U''}
\def\Svec{\vec{S}}
\def\mvec{\vec{m}}
\def\zdag{z^\dagger}
\def\sigmavec{\vec{\sigma}}
\def\sigvec{\sigmavec}
\def\svec{{\vec{n}\,}}
\def\mvec{\svec}
\def\lambdastar{\lambda_*}
\def\dim#1{\lbrack\!\lbrack #1 \rbrack\!\rbrack }
\def\chidagger{\chi^\dagger}
\def\nvec{\vec{n}}

\title{Quantum critical  spin liquids 
and conformal field theory 
in $2+1$ dimensions}
\author{Andr\'e  LeClair}
\affiliation{Newman Laboratory, Cornell University, Ithaca, NY} 
\date{December  2006}

\bigskip\bigskip\bigskip\bigskip

\begin{abstract}

We describe new conformal field  theories based on 
symplectic fermions that
can be extrapolated between $2$ and $4$ dimensions. 
The critical exponents depend continuously on the number
of components $N$ of the fermions and the dimension $D$.  
In the context of
anti-ferromagnetism,  the $N=2$ theory   is proposed to describe 
a deconfined quantum critical spin liquid corresponding to 
a transition between a N\'eel ordered phase and a VBS-like  phase.

\end{abstract}

\maketitle

\section{Introduction and summary of results}

This paper concerns some new renormalization group (RG) fixed points in 
$D=d+1=3$ dimensional quantum field theory  based on symplectic fermions.    
In $1d$ there exists a vast assortment of critical points 
described by $2D$ conformal field theory\cite{BPZ}.  
However 
in $2d$ the known critical theories are up to now  comparatively rare. 
Important examples are the 
Wilson-Fisher  fixed points which  are known to describe phase transitions 
in  classical statistical mechanics in $3$ spacial dimensions as a function
of temperature\cite{Wilson,Kogut}.   Further progress in some important
$2d$ condensed matter systems, in particular superconductivity
in the cuprates,   has been hindered by the lack of understood  critical points in
$3D$.  

This lengthy introduction will serve to summarize our main results. 
Since the motivation for our model initially came  from quantum
anti-ferromagnets in $2d$,  we begin by reviewing the aspects
of this problem that provide some perspective on our work. 
For a detailed  account of this subject with additional 
references to the original works  see Fradkin's book\cite{Fradkin}.

The Heisenberg hamiltonian for a collection of spins $\Svec_i$ on 
a $d$ dimensional lattice is
\beq
\label{I.1}
H = J \sum_{<i,j>} ~ \Svec_i  \cdot \Svec_j 
\eeq
For the anti-ferromagetic case ($J>0$), 
a continuum limit built over a staggered configuration close
to the N\'eel state leads to the euclidean action
\beq
\label{I.2}
S = \inv{2g} \int  d^D x  ~ \( \d_\mu \mvec \cdot \d_\mu \mvec \) ~~~+ 
S_\theta
\eeq
where $\mvec^2 = 1$ and $\d_\mu \d_\mu =  \sum_{\mu=1}^D  \d_{x_\mu}^2$.  
 The latter constraint makes it
an $O(3)$  non-linear sigma-model.   In any dimension, 
the topological term $S_\theta$ arises
directly in the map to the continuum when one formulates the
functional integral for the spins $\Svec_i$ using coherent 
states\cite{Wiegmann1,Fradkin88}, and is related to the
area swept out by the vector $\mvec$ on the 2-sphere.   In $1d$ it is given
explicitly by
\beq
\label{I.3}
S_\theta = \frac{\theta}{2\pi} \int d^2 x ~ \epsilon_{\mu\nu} 
\mvec \cdot (\d_\mu \mvec \times \d_\nu \mvec )
\eeq
with  $\theta = 2 \pi s$ where $s$ is the spin of $\Svec$:
$~~\Svec_i^2 = s(s+1)$. 
When $s$ is an integer,  $S_\theta$ has no effect in the
functional integral:   the model has  no non-trivial infra-red (IR) fixed
point and  the model is gapped.  For half-integer spin chains,
$S_{\theta = \pi}$ modifies the IR behavior and the model
has  an infra-red fixed point 
which necessarily has massless degrees of freedom.  This is 
the well-known Haldane conjecture\cite{Haldane}. 

The above understanding of  $1d$ spin chains relies to
a considerable extent on the Bethe-ansatz solution\cite{Bethe}, 
which shows massless excitations.   What is more subtle, and was only
realized much later,  is that the low-lying excitations 
are actually spin $1/2$ particles referred to as spinons\cite{Fadeev}.
These spinons are responsible for destroying the long-range
N\'eel order.  
The $\svec$ field is composed of two spinons, so this
is a $1d$ precedent to the $2d$ deconfined quantum critical points
discussed by Senthil et. al.\cite{Senthil}.  In the latter work,
numerous  arguments were given that in $2d$ there
should exist novel critical points that describe for example
the transition between N\'eel order and a valence-bond solid (VBS)
phase, 
and this idea strongly motivated our work initially.

\def\nvec{\svec}
\def\OM#1#2{O_#1^{(#2)}}
\def\Sp#1#2{Sp_{#1}^{(#2)}}

In $2d$ the term $S_\theta$ does not appear to have a significant role.
One way to  anticipate this is that unlike in $1d$ where 
$\mvec$ and $S_\theta$  are classically dimensionless (in
fact $S_\theta$ is exactly marginal), in $2d$  $\mvec$ has
classical dimension $1/2$ so that $S_\theta$ is already RG irrelevant 
before any quantum anomalous corrections.  

Remarkably in $2d$  a non-trivial IR  fixed point appears in the
Heisenberg anti-ferromagnet that doesn't rely on the existence of
$S_\theta$ and can be understood in the following way\cite{Halperin,Ye}.   
The non-linear constraint $\mvec^2 =1$ renders the non-linear
sigma model perturbatively non-renormalizable in $2d$. (Unlike in $1d$,
see \cite{Polyakov}.)
If a fixed point is understandable by Wilsonian RG,  this
non-renormalizability is potentially a serious problem.  However 
the infra-red  behavior is captured by the following 
scalar field theory:
\beq
\label{I.4}
S_{WF}  =  \int d^3 x \( \inv{2} \d_\mu \mvec \cdot \d_\mu \mvec  +~~
\tilde{\lambda} \,  (\mvec\cdot \mvec )^2 \)
\eeq
where now $\mvec$ is not constrained to be a unit vector
and hence is a linear sigma model but with interactions.   
The above model can be studied in the epsilon expansion around
$D = 4$ and the IR fixed point is seen perturbatively.  
(See for instance \cite{Peskin,Zinn}.) 
This is the universality class of the Wilson-Fisher (WF) fixed
point, even though it is a quantum critical point\cite{Sachdev}  at zero temperature. 
We will refer to this model as the $O(M)$ linear sigma model
and the WF fixed point conformal field theory as 
$\OM{M}{D}$ in $D<4$ dimensions.

A large part of the literature devoted to
the on-going search for other ground states of quantum spins
represents the $\mvec$ field in terms of spinon  fields $z$:
\beq
\label{I.5}
\mvec =  \zdag \vec{\sigma} z
\eeq
where $\sigmavec$ are the Pauli matrices and $z = (z_1 , z_2)=
\{ z_i  \} $ is a two component complex  bosonic spinor.
The constraint $\mvec^2 =1$ then follows from   the constraint
$\zdag z = 1$.   Coupling $z$ to a $U(1)$ gauge field $A_\mu$ 
with the covariant derivative $D_\mu = \d_\mu - i A_\mu$, then by 
eliminating the non-dynamical gauge field using it's equations of motion,
one can show that  the following actions are
equivalent:
\beq
\label{I.6}
\int d^{D} x ~  {\textstyle \inv{2}}
 \d_\mu \mvec \cdot \d_\mu \mvec = \int d^{D} x 
~| D_\mu z |^2  
\eeq
(One needs $\sigvec_{ij} \cdot 
\sigvec_{kl } = 2 \delta_{il}\delta_{jk}
- \delta_{ij} \delta_{kl}$.) 

Since the spinon fields have classical dimension $1/2$,
in terms of the spinon fields, the $(\svec\cdot\svec)^2$ term in
the action (\ref{I.4}) is a dimension $4$ operator which is irrelevant
in $2d$.  In analogy with the 
 WF fixed point, in order to deal with the non-renormalizability
it is 
natural to relax the constraint $\svec^2 = |z^\dagger z| =1$ 
and  to consider $(\zdag z)^2$ terms of dimension 2 which 
are RG relevant.  However  the fixed point is then still in the universality
class of the WF fixed point.  In an effort to perturb the WF fixed point,  
the authors in \cite{Senthil} 
 make the gauge field dynamical
by adding $(F_{\mu\nu})^2$ and thus consider a QED-like theory (in $3D$). 
It was proposed  that the model has a fixed point in the universality
class of a hedgehog suppressed $O(3)$ sigma model,   however
because of the  expected non-perturbative nature of the fixed point,
it hasn't been possible to compute any of it's critical exponents.   
It has  also proven difficult to see such a second-order transition in 
simulations of the model\cite{Proko}.

With the above background we now present the central idea of this paper.
The WF fixed point in this anti-ferromagnetic 
context is a quantum critical confined phase
since it is described in terms of the $\nvec$ degrees of freedom.  
A deconfined quantum critical point is defined then as 
one describable with the spinon $z$ degrees of freedom. 
We will postulate that 
 for a  deconfined critical point the 
 spinon field $z$ should actually be a fermion field,
henceforth denoted $\chi$,  and described by the action
\beq
\label{central.1}
S_\chi =  \int d^D x ~  \(  \d_\mu \chi^\dagger \d_\mu \chi  
  + \hat{\lambda} ~ |\chi^\dagger \chi|^2 \)
\eeq
where $\chi$ is a two-component complex field, $\chidag \chi = 
\sum_{i=1,2} \chidag_i \chi_i$.   The non-linear
constraint $\chidag \chi=1$ is obviously relaxed. 
Note that our  model contains no gauge field.  
In terms of real fields each component can be written
as $\chi = \eta_1 + i \eta_2$, $\chidagger = \eta_1 - i \eta_2$
and the free action is 
\beq
\label{central.2}
S_{\rm symplectic} = i \int d^D x  ~   \epsilon_{ij} \, \d_\mu \eta_i 
\d_\mu \eta_j 
\eeq
where $\ep_{12} = - \ep_{21} = 1$.  
The $2N$ real component version has the symmetry
$\eta \to U \eta$ where $U$ is a $2N\times 2N$ dimensional
matrix satisfying $U^T \epsilon_N U = \epsilon_N$ where
$\epsilon_N = \epsilon \otimes 1_N$.   This implies the theory 
has $Sp(2N)$ symmetry,  hence 
is sometimes referred to as a symplectic fermion.   This kind of theory 
 has a number of important applications in $1d$,
for instance to dense 
polymers\cite{Saleur},  and to disordered Dirac fermions 
in $2d$\cite{Guruswamy}.    It is known to be a non-unitary theory 
and this potential difficulty will be addressed below.

\def\kvec{{\vec{k}}}

Arguments suggesting that  symplectic fermions are natural in
this context are the following.  First of all,  in $1d$ the 
spinon is neither a boson nor a fermion but a semion, i.e. half
fermion as far as its statistics is concerned, so there is
a precedent for this kind of modified statistics.  We remind the reader that  there
is no spin-statistics theorem  for $2d$  relativistic theories since spin is not necessarily
quantized.  
Note also that the identity (\ref{I.6}) remains true if
$z$ is a fermion.     
Secondly, suppose the theory is asymptotically free in
the ultra-violet.  Then in this free, conformally invariant limit,
one would hope that the description in terms of $\svec$ 
or $ z$ are somehow equivalent, or at least have the same
numbers of degrees of freedom.   One way to count these degrees
of freedom is to study the theory at finite temperature $T=1/\beta$
and consider the free energy.   For a single species of free 
massless particle the free energy density is
\beq
\label{free}
\CF = \pm \inv{\beta} \int  \frac{d^d \kvec}{(2\pi)^d } ~ 
\log \( 1 \mp e^{-\beta\omega_\kvec} \) 
\eeq
where $\omega_\kvec = |\kvec|$ and the upper/lower sign
corresponds to bosons/fermions.     In $2d$
\beq
\label{free.2}
\CF = - c_3 \frac{\textstyle \zeta(3)}{2\pi} T^3
\eeq
where $c_3=1$ for a boson and $3/4$ for a fermion.    
(In $1d$ the analog of the above is $\CF = - c \pi T^2/6$, where 
$c$ is the Virasoro central charge\cite{Cardy,Affleck}.)  
Therefore one sees that the 3 bosonic degrees of freedom 
of an $\svec$ field  has the same $c_3$ as an $N=2$  component
$\chi$ field. This simple observation is what  first pointed us
in the direction of symplectic fermions.   It suggests a kind
of bosonization where 3 bosons are  equivalent to 4 fermions.  
In $1d$, one boson is equivalent to 2 fermions, and one
can explicitly construct the fermion fields in terms of 
bosons,  but we won't need to attempt the
analog here.  

Lastly,  and most importantly,  the symplectic fermion theory 
has an infra-red stable fixed point that is not in the WF universality class, and this is the main subject
of this paper.  The exponents can be computed in the very low
order epsilon expansion
around $D=4$ and they are in excellent agreement with the 
exponents found numerically for the hedgehog-free model studied by
Motrunich and Vishwanath\cite{Motrunich}.    
The agreement is
better than we anticipated.   We find 
\beq
\label{exps}
\eta = 3/4, ~~~\nu = 4/5, ~~~\beta = 7/10~~~~~~(N=2, D=3)
\eeq
 compared to
$\nu=.8 \pm 0.1,~~ \beta/\nu  = .85 \pm 0.05$.  The shift down to $3/4$ 
from the classical value $\eta = 1$ is entirely due to the
fermionic nature of the $\chi$ fields. 
Since we are discussing a zero temperature critical point,
temperature exponents are probably not important for comparison
with experiments, so    
we also define a magnetic exponent $\delta$ so
that $\langle \nvec \rangle \sim B^{1/\delta}$ where
$B$ is an external magnetic field and find $\delta = 17/7$.  
It is the fermionic statistics of $\chi$ that is also 
ultimately responsible for the largeness of these exponents
in comparison with the bosonic WF fixed point. 
We thus conjecture that our 2-component model
describes the fixed point in \cite{Motrunich}, which is
believed to be a deconfined critical point of the kind
discussed by Senthil et. al.\cite{Senthil}.     It is
important to point out that there are no 
``emergent photons'' in our model.

Let us return now to the issue of topological terms.  So far
we have ignored the gauge field $A_\mu$ in the action (\ref{I.6}).
Wilczek and Zee have shown  how to include a topological term\cite{WZee}
that is intrinsic to $2d$, which is a Chern-Simons or Hopf term:
\beq
\label{I.7}
S_{\rm CS} = \frac{\vartheta}{8 \pi^2} \int d^3 x  ~ 
\epsilon_{\mu\nu\lambda} A^\mu F^{\nu\lambda}
\eeq
It is well-known that $\vartheta$ modifies the statistics of
the $z$,  and this provides an obvious  mechanism for obtaining
the fermionic statistics of the $\chi$ field when $\vartheta = \pi$.  
It should be emphasized that this Chern-Simons term has
nothing to do with the $S_\theta$ term discussed above.  
Whereas the latter arises directly in the map to the continuum,
the coefficient of the Chern-Simons term is more subtle.   
The  consensus is  that for a square lattice 
$\vartheta =0$\cite{Fradkin88,Haldane88,Wen88,Ioffe88,Dombre88}.  

The idea that the fermionic $\chi$ model should correspond 
to the action (\ref{I.6}) with the addition of a Chern-Simons term 
with $\vartheta = \pi$ helps to resolve the problem that 
the symplectic fermion theory is non-unitary,  since the 
Chern-Simons description is thought to be unitary.   In this
description the $\chi$ particles are to be viewed as having
$\pi$ flux attached microscopically.   We return to the issue of
the non-unitarity in section IV  and give a different possible resolution of
it based on a simple projection of the Hilbert space onto pairs of particles.

In this paper we do not address what kind of microscopic
theory can give  rise to $\vartheta\neq 0$ and whether it is related
for instance to the spin $s$ of the $\Svec$, as in
$\theta = 2\pi s$.   But we can nevertheless
state  a $2d$ version of the Haldane conjecture:  
When $\theta=0 ~({\rm mod ~ 2\pi})$, the quantum critical point
is confined and in the universality class of the WF
fixed point.   On the other hand when $\theta = \pi ~({\rm mod ~ 2\pi})$, 
the quantum critical point is in the different universality class of
the fixed point in the $N=2$ component fermionic theory described 
in this  paper.

The bulk of this paper analyzes the RG for the
symplectic model to support the above statements. 
In section II we compute
the beta function in $D=4$ from the effective potential,
and display the fixed point in $D<4$.   In section III 
we define the critical exponents and relate them to
anomalous dimensions in the symplectic fermion theory.  
The relevant correlation functions are computed to lowest
orders  in position space and this determines the exponents 
as a function of $N$ and $D$.     
(In terms of Feynman diagrams, which we manage to avoid,  this involves 
one and two loop diagrams. For the exponent $\nu$ one actually  only
needs 1-loop results.)

In section IV we study our model in a hamiltonian framework in momentum
space and show how the non-unitarity in manifested.    We argue that
a simple projection onto pairs of particles renders the Hilbert space unitary.
We also discuss possible applications to superconductivity in the cuprates.

\section{RG beta function}

\subsection{Functional RG}

The 3D fixed point can studied systematically in the
epsilon expansion around $D=4$\cite{Peskin,Zinn}. 
The Feynman diagram techniques developed for the WF fixed point, 
namely dimensional regularization, lagrangian counterterms, etc, 
is easily generalized to our fermionic theory. 
  However since we
will work to lowest orders only,  we need not develop
the epsilon expansion in much  detail since the
required quantities can be computed directly in $4D$.  
Working with  position space correlation functions
also   helps to clarify the 
physical content. 

For the beta function it turns out to be convenient to
work with the Coleman-Weinberg 
effective potential.  This avoids Feynman diagrams
and  can be especially
useful if the potential has complicated anisotropy, though
this will not be investigated here.  It also helps to
keep track of the all-important fermionic minus signs.  

Let $\chi$ denote a vector of complex fermionic (Grassman)  fields 
$\chi = (\chi_1 , \ldots , \chi_N) = \{ \chi_i \}$,
and consider the euclidean action:
\beq
\label{FRG.1}
S_\chi  = \int d^D x \(  2  \d_\mu \chidag \d_\mu \chi  + U (\chi^\dagger, \chi) \)
\eeq
where $U$ is the potential. 
For the purposes of Grassman 
functional integration,  
it is convenient to arrange $\chi, \chidag$ into a $2N$ dimensional vector
$\Psi = (\chi , \chidag)$.   
Note that $\Psi^\dagger = \Psi^T \Sigma_1$ where
in block form $\Sigma_1 = \( \matrix{0&1\cr 1&0\cr} \)$, so that
$\Psi^\dagger$ is not independent of $\Psi$ as far as functional
integration is concerned.

Consider  the functional integral
\beq
\label{FRG.2}
Z = \int D\Psi ~ e^{- S}
\eeq
in  a saddle point approximation. 
Expanding $\Psi$ around $\Psi_0$ and performing the fermionic 
gaussian functional 
using the formula
\beq
\label{FRG.3}
\int D\Psi ~ e^{- \Psi^\dagger A \Psi } = \sqrt{\Det A} 
\eeq 
one obtains  $Z \sim e^{-S_{\rm eff}}$
where the effective action is 
\beq
\label{FRG.4}
S_{\rm eff} = S(\Psi_0) - \inv{2} \Tr \log A 
\eeq
and the Trace is  functional.
The operator $A$ is
\beq
\label{FRG.4b}
A = {\scriptstyle  \( \matrix{ - \d^2 &0\cr 0 &\d^2 \cr } \)} - \inv{2} \Upp
\eeq
where
 $\Upp$ is a matrix of second derivatives:
\beq
\label{FRG.8b}
\Upp = \( 
\matrix{ \frac{ \d^2 U}{\d \chidag \d \chi} &
\frac{ \d^2 U}{\d \chidag \d \chidag}\cr
\frac{ \d^2 U}{\d\chi  \d \chi}&
\frac{ \d^2 U}{\d \chi \d \chidag}\cr}
\)
\eeq
Assuming $\Upp$ is constant, 
the trace can be computed in  momentum space. Dividing by the volume one
defines the effective potential
\beq
\label{FRG.5}
V_{\rm eff} = U - \inv{2} \tr  \int \frac{d^D k}{(2\pi)^D} ~ 
\log \( {\scriptstyle \( \matrix{k^2 & 0\cr 0&-k^2 \cr} \)} - \inv{2} \Upp \)
\eeq

Since we are interested  only in the term that determines the beta function
in $D=4$,
we expand the $\log$ to second order in $\Upp$ which involves 
$1/k^4$.   Performing the integral over $k$ with an ultra-violet
cut off $\mu$:
\beq
\label{FRG.6}
\int^\mu  
\frac{d^4 k}{(2\pi)^4} \inv{k^4} = \frac{\log \mu}{8\pi^2} 
\eeq
one finds
\beq
\label{FRG.7}
V_{\rm eff} = U + \frac{\log\mu}{128 \pi^2} ~  
\tr ( \Sigma_3 \Upp \Sigma_3 \Upp ) 
\eeq
where in block form
\beq
\label{FRG.8}
\Sigma_3 = \( \matrix{1&0\cr0&-1\cr } \)
\eeq

The  renormalization group  of the   potential 
then follows from requiring
\beq
\label{FRG.9}
V_{\rm eff} (U, \mu ) = V_{\rm eff} (U(a), e^a \mu )
\eeq
We adopt the usual convention in statistical physics 
and define the beta function as the flow toward the infra-red,
i.e. to low energies:  $d U / d \ell = - d U / d a$ 
where $e^\ell$ is a length scale. 
Finally one gets the beta function:
\beq
\label{FRG.10}
  \frac{ d U}{d \ell}  = \inv{128 \pi^2}   
\tr ( \Sigma_3 \Upp \Sigma_3 \Upp )
\eeq
The flow toward the IR then  corresponds to increasing $\ell$. 

The RG flow equation (\ref{FRG.10}) is a functional RG since
it determines the flow of the {\it entire} potential.   
It is then clear that not all potentials are renormalizable:
only if $\tr (\Sigma_3 \Upp \Sigma_3 \Upp)$ is proportional
to $U$ is the theory closed under RG flows.   If additional terms 
are generated in the trace, they must be included in $U$
until one obtains something renormalizable.  By ``renormalizable''
we here mean that the flow just amounts to the flow
of some couplings.

\def\chidag{\chi^\dagger}

\subsection{Beta function and fixed point}

Let us take the following normalization of the coupling $\lambda$
\beq
\label{fp.1}
U_\lambda  = 16 \pi^2 \lambda ~ U , ~~~~~U\equiv |\chidag\chi|^2 
\eeq
Denoting $\d_i = \d / \d \chi_i $, $\d_i^\dagger = \d/\d \chidag$ one
has 
\barray
\label{fp.2}
\d_j \d_i U &=& -2 \chidag_i \chidag_j 
\\ \nonumber
\d_j^\dagger \d_i U &=& -2 \delta_{ij} \chidag \chi + 2 \chidag_i \chi_j 
\earray
Evaluating the trace in eq. (\ref{FRG.10}) one finds that 
it is proportional to $U$ since $U$ was chosen to be isotropic. 
We also need that the classical dimension of 
$\chi$ is $(D-2)/2$ which implies
the dimension of $\lambda$ is $4-D$. Since the leading linear term
always has a slope equal to the
 classical dimension of $\lambda$, the beta function is 
\beq
\label{fp.3}
\frac{d\lambda}{d\ell} = (4-D) \lambda + (N-4) \lambda^2 
\eeq

The above beta function has a zero at 
\beq
\label{fp.4}
\lambdastar = \frac{4-D}{4-N}
\eeq
Note that $\lambda_*$ changes sign at $N=4$.  
It is not necessarily a problem to have a fixed point at
negative $\lambda$ since the particles are fermionic:
the energy is not unbounded from below because of the
Fermi sea.   Near $\lambda_*$ one has that
$d \lambda /d\ell \sim (D-4)(\lambda - \lambdastar)$
which implies the fixed point is IR stable regardless of  
the sign of $\lambda_*$,  so long as $D<4$. When 
$D>4$ we have a short distance fixed point that is not
asymptotically free.

\def\xvec{{\bf x}}
\def\yvec{{\bf y}}
\def\zvec{{ \bf z}}
\def\gammachi{{\gamma_\chi}}
\def\gammam{{\gamma_m}}
\def\absx{|\xvec|}

\section{Scaling theory and critical exponents}

There are two aspects of the scaling theory
that differ from the usual WF fixed point
for classical phase transitions.  The first is 
that $\nvec$ is a composite field in terms of the
$\chi$'s.    The second
is that  due to the fermionic nature of $\chi$, 
some of the exponents are in fact {\it negative}.

\subsection{Definition of the exponents for the $\nvec$ field }

Though the spinons $\chi$ are deconfined, it
is still physically meaningful to define exponents
in terms of the original order parameter $\nvec$,
which is represented by
\beq
\label{scale.0}
\nvec = \chi^\dagger \vec{\sigma} \chi
\eeq
When $N=2$,  $\vec{\sigma}$ are the Pauli matrices; 
for general $N$ they should be taken as representations
of the Clifford algebra appropriate to spinor representations of $O(N)$,
however we will not need these details in this paper.  
We then define the exponent $\eta$ as the one
characterizing the spin-spin correlation function:
\beq
\label{scale.1}
\langle \svec (\xvec ) \cdot \svec (0) \rangle \sim 
\inv{|\xvec|^{D-2 + \eta}}
\eeq
For the other exponents we need a measure of the
departure from the critical point;  these are the
parameters that are tuned to the critical point in
simulations and experiments:
\beq
\label{scale.2}
S_\chi \to S_\chi + \int d^D x ~(m^2 \, \chidag \chi  + 
\vec{B}\cdot \nvec  )
\eeq
Above, $m$ is a mass and $\vec{B}$ the magnetic field.
For classical temperature phase transitions, typically 
$m^2\propto (T-T_c)$.  
The correlation length exponent $\nu$,   and magnetization 
exponents $\beta, \delta$ 
are then defined by
\beq
\label{scale.3}
\xi \sim m^{-\nu}, ~~~~~\langle \svec \rangle \sim m^\beta \sim 
B^{1/\delta} 
\eeq
Above $\langle \nvec \rangle$ is the one-point function of
the field $\nvec ({\bf x})$ and is independent of ${\bf x}$ 
by the assumed translation invariance.

To  streamline the discussion,  let $\dim{X}$ denote the scaling
dimension of $X$ in energy units,  including the non-anomalous 
classical contribution which depends on $D$.  
An action $S$ necessarily
has $\dim{S}$ =0.  Using $\dim{d^D \xvec} = -D$,  the
classical dimensions of couplings and fields are
determined from $\dim{S}=0$.  
The above exponents are then functions of $\dim{\nvec}$
and $\dim{m}$.     Since $\dim{\xi} = -1$,  one has $\nu = - \dim{\xi}/\dim{m} = 1/\dim{m}$.   
This, together with eq. (\ref{scale.1}), implies
\beq
\label{etanu} 
\eta = 2 \dim{\nvec}   + 2 -D, ~~~~~\nu = 1/ \dim{m} 
\eeq
One also has
\beq
\label{betanew}  
\beta =  \dim{\nvec}/ \dim{m} = \frac{\nu}{2} (D-2 + \eta) 
\eeq

The magnetic exponent is treated similarly. 
Treating $\vec{B}$ as a coupling, then
$\dim{\vec{B}} + \dim \nvec  = D$,  which implies
\beq
\label{scale.8b}
\delta = \frac{ D-\dim{\nvec}  }{\dim{\nvec} } = 
\frac{ D + 2 - \eta }{D + \eta -2 }
\eeq

\subsection{Relation to $\chi$ field exponents}

The fundamental exponents of the symplectic fermion  theory 
are the anomalous dimensions $\gammachi, \gammam$ of $\chi$ and $m$,  defined as: 
\beq
\label{scale.4}
\dim\chi \equiv  (D-2)/2 + \gammachi ,  ~~~~~\dim{m} \equiv 1-\gammam
\eeq
The $\gammachi$ exponent determines the two point function 
of the $\chi$ fields:
\beq
\label{scale.5}
\langle \chidag (\xvec ) \chi (0) \rangle \sim 
\inv{ \absx^{D-2 + 2 \gammachi}} ,
\eeq
whereas $\nu$ is completely determined by $\gammam$:
\beq
\label{nunew}
\inv{\nu} = 1- \gammam
\eeq

The  scaling dimension 
$\dim{\nvec}$ is not a simple function of $\gammachi$ to all orders since  $\nvec$ 
must be treated as a composite operator.   However, since $\nvec$ is quadratic
in $\chi$,   let us  assume that to lowest order
$\dim{\nvec} = 2\dim{\chi}$.  
With this assumption one has 
\beq
\label{scale.7}
\eta = D-2 + 4 \gammachi
\eeq

\subsection{Computation of correlation functions}

To compute $\gammachi, \gammam$ we need to consider
the following correlation functions in $4D$.    Introduce
an ultra-violet cut-off $\mu$ as follows 
$\int d^4 \xvec \to \int_{1/\mu} d^4 \xvec$. 
Suppose the one-point function of $\chi^\dagger \chi$ satisfies 
\beq
\label{scale.10}
\langle \chi^\dagger \chi \rangle \sim m^2 ( 1-2\alpha \log \mu ) 
\approx (m \mu^{-\alpha})^2
\eeq
Then the ultra-violet divergence is canceled by letting
$m\to m(\mu) = m \mu^{\alpha}$.   This implies  
\beq
\label{scale.11}
\gammam = \alpha 
\eeq
This is equivalent to the statement $\dim{\chidag \chi } = D-2 + 2 \gammam$.

The exponent $\gammachi$ is determined from the  two-point function.
Suppose 
\beq
\label{scale.12}
\langle \chidag_i (\xvec ) \chi_j (0) \rangle \sim 
\frac{\delta_{ij}}{\absx^2} ( 1- 2\alpha' \log \absx ) 
\approx \frac{\delta_{ij}}{ \absx^{2 + 2\alpha'} }
\eeq
Then 
\beq
\label{scale.13}
\gammachi = \alpha' 
\eeq

We now describe the lowest order contributions to
the needed  correlation functions.
 One needs the
free 2-point functions:
\beq
\label{corr.1}
\langle \chidag_i (\xvec ) \chi_j (0) \rangle = - 
\langle \chi_i (\xvec ) \chidag_j (0) \rangle 
= - \frac{ \delta_{ij} }{8 \pi^2 \absx^2 }  ~~~~~({\rm when} ~ \lambda =0)
\eeq

Expanding the functional integral perturbatively in $\lambda$,
to first order one has:
\beq
\label{corr.2}
\langle \chidag \chi \rangle_\lambda = 
\langle \chidag \chi \rangle_0 - 16 \pi^2 \lambda \int d^4 \yvec ~ 
\langle \chidag \chi (0) U(\yvec ))\rangle  
\eeq
Using Wick's theorem, the integrand goes as $1/|\yvec|^4$ 
and the $y$-integration gives $\log \mu$:
\beq
\label{corr.2b}
\langle \chidag \chi \rangle_\lambda = \Bigl( 1- \lambda (1-N) \log \mu \Bigr) 
\langle \chidag \chi \rangle_0
\eeq
From dimensional analysis $\langle \chidag \chi \rangle_{\lambda =0}
\propto m^2$.  Thus
\beq
\label{corr.3}
\gammam = \frac{ \lambda (1-N)}{2}
\eeq
The negative sign in  $1-N$ in $\gammam$ is entirely statistical in origin. 

To compute $\gammachi$ we need to go to second order:
\barray
\nonumber
\langle \chidag_i (\xvec ) \chi_j (0)\rangle_{\lambda^2} &=&
{\textstyle \inv{2}} (16 \pi^2 \lambda)^2 \int d^4 \yvec \int d^4 \zvec ~ 
\langle \chidag_i (\xvec ) \chi_j (0) \, U(\yvec )\,  U(\zvec) \rangle 
\\ 
\label{corr.4}
&=& \delta_{ij} \frac{ (1-N)\lambda^2 }{32 \pi^6} ~ I(\absx)
\earray
where $I(\absx )$ is the integral:
\beq
\label{corr.5}
I(\absx) = \int d^4 \yvec \int d^4 \zvec ~
\inv{ |\yvec|^2 |\xvec - \zvec|^2 |\yvec - \zvec|^6 }
\eeq
It is evaluated in Appendix A
where it is shown:
\beq
\label{corr.6}
I(\absx ) = \frac{ 2 \pi^4 \log \absx}{\absx^2}
\eeq
This in turn implies: 
\beq
\label{corr.7}
\gammachi = \frac{(1-N)\lambda^2}{4}
\eeq

\subsection{Values of exponents  for 
arbitrary $N,D$. }

At the fixed point, one substitutes $\lambda = \lambda_*$ into
the above expressions.  The $\chi$ field exponents are:
\beq
\label{values.1}
\gammam = \frac{(4-D)(1-N)}{2(4-N)}, ~~~~~
\gammachi = \frac{(4-D)^2 (1-N) }{4 (4-N)^2 }
\eeq
which in turn imply the following $\nvec$ exponents:
\barray
\label{values.2}
\nu &=& \frac{2(4-N)}{(2-D)N + D +4} 
\\ \nonumber
\beta &=& \frac{ 2(D-2)(N^2 - 4N +12) + D^2 (1-N) }{(4-N)(D(1-N) + 2(N+2) )}
\earray
The $\eta, \delta$ exponents are given in terms of $\beta,\nu$ 
in eqs. (\ref{betanew},\ref{scale.8b}).  

For $N=2$,  the above expressions give the results quoted in the 
introduction:
$\nu = 4/5, \beta = 7/10,  \eta = 3/4$ and $\delta = 17/7$.

\section{Hamiltonian description and unitary projection}

In this section we give a hamiltonian description of our
model in momentum space and address the non-unitarity issue in
a  different way than in the Introduction. 
We also suggest how 
 our work may  provide 
some clues toward the  
understanding of superconductivity in the cuprates, 
which is believed to be a $2+1$ dimensional problem\cite{Anderson2}.  
To do this, one must turn  to the language of the Hubbard
model.   In the anti-ferromagnetic phase of the Hubbard model,
the spin field $\nvec = c^\dagger \sigmavec c $,  where
$c$ are the physical electrons.   Therefore in applying our
model to the Hubbard model,  the symplectic fermion $\chi$ 
is a descendant of  the electron, so it seems it can  carry electric charge.  
Consider the zero temperature phase diagram of the cuprates as a function of
the density of holes.   At low density there is an anti-ferromagnetic
phase.  Suppose that  the first quantum critical point
is a transition from a N\'eel ordered 
 to a VBS-like  phase and is well described by our
symplectic fermion model at $N=2$.  Compelling
evidence for a VBS like phase has recently been
seen by Davis' group\cite{Seamus};  and it in fact resembles
more a ``VBS spin glass''.   The superconducting
phase actually originates from this VBS-like phase.    
Beyond the first deconfined quantum  critical point
at higher density,  it is then possible that the 2-component 
$\chi$ fields capture
the correct degrees of freedom for the description of this 
VBS-like phase.    It is important to point out that we are
imagining that the spinons are still deconfined in the VBS-like phase,
in contrast to ideas in \cite{Senthil}.    These fermionic spinon quasi-particles
acquire a gap away from the critical point,
which is described by the mass term in
eq. (\ref{scale.2}).  Note that away from the quantum
critical point, the particles already have a gap $m$
because of the relativistic nature of the symplectic
fermion. 

Particles with a relativistic kinetic energy are actually not
entirely new  in $2d$ physics:  graphene appears to have
massless particles described by the Dirac equation\cite{Geim,Kim}, rather
than the symplectic fermion theory we considered. 
In graphene the origin of the massless Dirac equation is
simply band theory on a hexagonal lattice, so the origin
of massless relativistic particles is entirely different (and much simpler) 
than
the origin of our symplectic fermions.

\def\adag{a^\dagger}
\def\bdag{b^\dagger} 
\def\kvec{{\bf k}}

The hamiltonian of the symplectic
fermion is
\beq
\label{highT.1}
H = \int d^2 \xvec  \(  2 \d_t \chi^\dagger \d_t \chi + 
2 \vec{\nabla} \chi^\dagger \cdot \vec{\nabla} \chi 
+ m^2 \chidag \chi  +  \tilde{\lambda} ~ (\chidag \chi)^2 \)
\eeq
Expand the field in terms of creation/annihilation operators
as follows
\beq
\label{highT.2}
\chi (\xvec ) = \int \frac{ d^2 \kvec }{4\pi \sqrt{\omega_{\kvec}}} 
\( a_\kvec \, e^{-i\kvec \cdot \xvec} + b_\kvec \, e^{i \kvec\cdot \xvec} \)
\eeq
and similarly for $\chidag$, where $\omega_\kvec = \sqrt{\kvec^2 + m^2 }$. 
Canonical quantization of the $\chi$-fields leads to
\beq
\label{hight.3}
\{ \bdag_\kvec , b_{\kvec'} \} = - \{ \adag_\kvec , a_{\kvec'} \} = 
\delta_{\kvec , \kvec'} 
\eeq
The free hamiltonian is then
\beq
\label{highT.4}
H_0 = \int d^2 \kvec ~  \omega_\kvec   \, 
\( \adag_\kvec a_\kvec + \bdag_\kvec b_\kvec \) 
\eeq
Because of the minus sign in eq. (\ref{hight.3}) this is a two-band
theory:
\barray
\label{highT.5}
H_0 ~ \bdag_\kvec |0\rangle &=& \omega_\kvec 
~\bdag_\kvec |0\rangle 
\\ \nonumber
H_0 ~ \adag_\kvec |0\rangle &=& - \omega_\kvec 
~ \adag_\kvec |0\rangle 
\earray
A two-band structure of this kind has been observed experimentally \cite{Seamus}
and in the Hubbard model\cite{Scalapino}.  

The unconventional minus sign in the anti-commutator of
the $a,  \adag$ is a manifestation of the non-unitarity.   
In particular it implies that  the state $|\kvec \rangle_a  = \adag_\kvec | 0 \rangle $
has negative norm:  ${}_a \langle \kvec | \kvec' \rangle_a = - \delta_{\kvec \kvec'}$.  
On the other hand,  the two-particle state $|\kvec_1 , \kvec_2 \rangle_a$ 
has positive norm.     We thus propose to resolve the non-unitarity 
problem by simply projecting the Hilbert space onto even numbers
of $a$-particles.   It is clear that if they arise from deconfinement of the
$\nvec$ field,  they will be created in pairs.

Let us turn to the interactions.  
In the VBS-like  phase the $\chi$-particles are charged
fermions and it's possible that additional phonon interactions,
or even the $\chi^4$ interactions that led to the critical theory,
could
lead to a pairing interaction that causes them to condense
into Cooper pairs just as in the usual BCS theory.   
Recent numerical work on the Hubbard model suggests that
the Hubbard interactions themselves can provide
a pairing mechanism\cite{Scalapino}.  
 The $(\chidag \chi)^2$ interaction
is very short ranged since it corresponds to a $\delta$-function 
potential in position space.  Because of the relativistic nature
of the fields, the interaction gives rise to a variety of 
pairing interactions.   Let us examine pairing interactions
within each band that resemble BCS pairing.
  If  all momenta have roughly the same magnitude
$|\kvec|$,  then the interaction gives the term (up to factors of $\pi$):
\beq
\label{highT.6}
H_{\rm int} =   - \tilde{\lambda} \sum_{\kvec; ~i,j = \uparrow, \downarrow} 
\( \adag_{\kvec , i} \adag_{-\kvec , j} a_{-\kvec , i} a_{\kvec , j} 
 + (a\to b) \) + ....
\eeq
The overall minus sign of the interaction is due to the fermionic
statistics.   To compare with the BCS theory,  the interaction
contains terms such as
\beq
\label{highT.7}
H_{\rm int} = - \tilde{\lambda} \sum_\kvec \(  
\adag_{\kvec\uparrow} \adag_{-\kvec \downarrow} a_{-\kvec \uparrow}
a_{\kvec \downarrow} \) +....
\eeq
Because of the overall minus sign this is an attractive 
pairing interaction as in BCS.  One difference is that
some pairing interactions have the spins flipped in comparison with
BCS.

This discussion has been expanded and  a few more results derived
in \cite{AndreTc}.  

\section{Conclusions}

We have described new  $3D$ fixed points based on symplectic fermions 
and have applied the $N=2$ case to deconfined quantum criticality.     
In this interpretation,  the symplectic fermions are the deconfined spinons.  
The main evidence for our model is the agreement with the exponents
found in \cite{Motrunich} for the hedgehog-free $O(3)$ sigma model.   
The fermionic nature of the $\chi$-field  is what is responsible for the
negative contributions to exponents like $\eta$ that bring it below the
classical value $\eta = 1$.   

Our model is non-unitary,  however we have addressed this in two ways,
one based on a  unitary Chern-Simons description,  the other based on
a projection of the Hilbert space onto even numbers of particles.

After this work was completed,  new simulations by Sandvik report evidence
for a deconfined quantum critical point in a Heisenberg model with four-spin
interactions.    There it was found that $\nu = 0.78 \pm 0.03$ consistent with
\cite{Motrunich} and with our  prediction of $\nu = 4/5$.   The $\eta$ exponent
on the other hand,  $\eta = 0.26\pm 0.03$ is quite different from both our
result and  the one in \cite{Motrunich}.    This could simply mean that the 
critical point in the four-spin interaction model is in a different universality class.
It could also mean there are significant corrections to our exponents at higher order or due
to the compositeness of $\nvec$;
recall the fixed point at lowest order occurs at the relatively large value $\lambda_* = 1/2$.    
Higher order computations are currently in progress and we hope to report our
results in the near future.

Our exponents are well defined for any $N<4$,   including negative $N$.   Based on
the comparison of exponents,  our  model for $N$ a negative integer was proposed 
to describe $O(M)$ models such
as the Ising model in \cite{AndreIsing} 
with the identification  $M=-N$.

\section{Acknowledgments}

I would like to thank   S. Davis,  C. Henley,  P.-T. How,  A. Ludwig, 
E.  Mueller,  M. Neubert,   F. Noguiera,  N. Read,
S. Sachdev,  A. Sandvik, T. Senthil,  J. Sethna  and Germ\'an Sierra  for discussions.

\section{Appendix A}

In this appendix we do the integral $I(\absx)$ in eq. (\ref{corr.5}).
Shifting $\zvec = \zvec' + \yvec$, 
 and using  the identity
\beq
\label{A.1}
\inv{AB} = \int_0^1 dt ~ 
\inv{ (tA + (1-t)B)^2 }
\eeq
one can do the integral over $\yvec$:
\beq
\label{A.2}
\int_0^L d^4 \yvec ~ \inv{ |\yvec|^2 |\yvec - {\bf w}|^2 }
= \pi^2 \( \log( |{\bf w}|^2/L^2 ) - 3 \)
\eeq
where ${\bf w}  = \xvec - \zvec'$ and $L$ is an IR cut-off. 
Introducing an ultra-violet cut-off $\mu$,  
$\int d^4 \xvec \to \int_{\mu^{-1}}^L d^4 \xvec $
the integral over $\zvec'$ can be performed  giving  
\beq
\label{A.3}
I(\absx ) = \frac{\pi^4}{\absx^2} \(  2 \log (\absx/L ) + 1 \)  
- 6 \pi^4 \mu^2 
\eeq

\vspace{.4cm}

\begin{thebibliography}{99}

\bibitem{BPZ}   A. A. Belavin, A. M. Polyakov and A. B. 
Zamolodchikov, Nucl. Phys. {\bf B241} (1984) 333. 

\bibitem{Wilson} K. G. Wilson and M. E. Fisher, 
Phys. Rev. Lett. {\bf 28} (1972) 240. 

\bibitem{Kogut}  K. G. Wilson and J. Kogut, Phys. Rep. {\bf 12} 
(1974) 75. 

\bibitem{Fradkin}   E. Fradkin, {\it Field Theories of Condensed
Matter Systems}, Frontiers in physics vol. 82, Addison-Wesley 
1991. 



\bibitem{Wiegmann1}  P. Wiegmann, 
Phys. Rev. Lett. {\bf 60} (1988) 821. 

\bibitem{Fradkin88}  E. Fradkin and M. Stone, 
Phys. Rev. {\bf B38} (1988) 7215.


\bibitem{Haldane}   F. D. M. Haldane,
Phys. Lett. {\bf 93A} (1983) 464;  Phys. Rev. Lett. {\bf 50} 
(1983) 1153. 


\bibitem{Bethe}  H. Bethe, Zeitschrift f\"ur Physik {\bf 71} 
(1931) 205.

\bibitem{Fadeev}  L. D. Faddeev and L. H. Takhtajan, 
Phys. Lett. {\bf 85A} (1981) 375.

\bibitem{Senthil}  T. Senthil, L. Balents, S. Sachdev, 
A. Vishwanath and M. P. A. Fisher,
Phys. Rev. {\bf B70} (2004) 144407,
 cond-mat/0312617

\bibitem{Halperin}  S. Chakravarty, B. I. Halperin and D. R. Nelson,
Phys. Rev. {\bf B39} (1989) 2344. 

\bibitem{Ye}   A. V. Chubukov, S. Sachdev and J. Ye,
 Phys.  Rev.  {\bf B49} (1994) 11919
cond-mat/9304046

\bibitem{Polyakov}  A. M. Polyakov, {\it Gauge Fields and Strings},
Contemporary Concepts in Physics vol.3, Harwood 1987. 

\bibitem{Peskin}   M. E. Peskin and D. V. Schroeder, 
{\it An Introduction to Quantum Field Theory},  Addison-Wesley 1995.

\bibitem{Zinn}   J. Zinn-Justin,  
{\it Quantum Field Theory and Critical Phenomena}, 
2nd ed.,  Oxford Univ. Press 1993. 


\bibitem{Sachdev}   S. Sachdev, {\it Quantum Phase transitions},
Cambridge University Press, 1999.


\bibitem{Proko}    A. B. Kuklov,  N. V. Prokof'ev,  B. V. Svistunov, and
M. Troyer,  cond-mat/0602466.


 \bibitem{Saleur}  H. Saleur, Nucl. Phys. {\bf B382} (1992) 486,
hep-th/9111007. 

\bibitem{Guruswamy} S. Guruswamy, A. LeClair and A. W. W. Ludwig, 
Nucl.Phys. {\bf B583}  (2000) 475. 




\bibitem{Cardy}  J. Cardy,  J. Phys. {\bf A17} (1984) 385.

\bibitem{Affleck} I. Affleck,  Phys. Rev. Lett. {\bf 56} (1986) 746. 

\bibitem{Motrunich}  O. I. Motrunich and A. Vishwanath, 
Phys. Rev. {\bf B70} (2004) 075104, 
cond-mat/0311222. 







\bibitem{WZee}  F. Wilczek and A. Zee, 
  Phys. Rev. Lett. {\bf 51} (1983) 2250. 


\bibitem{Haldane88}  F.D.M. Haldane,  Phys. Rev. Lett. {\bf 61}
(1988) 1029.

\bibitem{Wen88}   X. G. Wen and A. Zee,  Phys. Rev. Lett. {\bf 61}
(1988) 1025. 

\bibitem{Ioffe88}  L. Ioffe and A. Larkin, 
Int. Jour. Mod. Phys. {\bf B2} (1988) 203.

\bibitem{Dombre88} D. Dombre and N. Read,
Phys. Rev. {\bf B38} (1988) 7181






\bibitem{WuZee} Y.-S. Wu and A. Zee,  Phys. Lett. {\bf 147B}
(1984) 325. 

\bibitem{Geim}  K. S. Novoselov, A. K. Geim,  S. V. Morozov,
D. Jiang,  M. I. Katsnelson, I.V Grigorieva, S. V. Dubonos
and A. A. Firsov, Nature {\bf 438} (2005) 197. 

\bibitem{Kim}  Y. Zhang,  Y.-W. Tan, H. L. Stormer and P. Kim, 
Nature {\bf 438} (2005) 201. 


\bibitem{Anderson2}  P. W. Anderson, Science {\bf 235} (1987) 1196.

\bibitem{Seamus}  Private discussions with Seamus
Davis;  Y. Kohsaka et. al.,  submitted (2006)



\bibitem{Scalapino}  T. A. Maier, M. S. Jarrell and D. J. Scalapino,
cond-mat/0606003, 0608507.  


\bibitem{AndreTc}  A. LeClair,  {\it Quantum critical spin
liquids and superconductivity in the cuprates},  cond-mat/0610816



\bibitem{Sandvik} A. W. Sandvik, cond-mat/0611343 



\bibitem{AndreIsing}  A. LeClair,  {\it $3D$ Ising and other models
from symplectic fermions,}   cond-mat/0610817





\end{thebibliography}
\end{document}